\documentclass{article}
\usepackage{spconf,amsmath,graphicx}
\usepackage{caption}
\usepackage{subcaption}
\usepackage{booktabs}
\usepackage[inline]{enumitem}
\usepackage{todonotes}
\usepackage{paralist}

\PassOptionsToPackage{hyphens}{url}\usepackage[hidelinks]{hyperref}
\title{Invisible Threats: Backdoor Attack in OCR Systems}
\name{Mauro Conti$^{1, 2}$, Nicola Farronato$^{1}$, Stefanos Koffas$^{2}$, Luca Pajola$^{1}$, Stjepan Picek$^{3,2}$}
\address{
    $^{1}$ University of Padua, Italy \\
    $^{2}$ Delft University of Technology, The Netherlands \\
    $^{3}$ Radboud University, The Netherlands
}

\begin{document}
%
\maketitle
\begin{abstract}
Optical Character Recognition (OCR) is a widely used tool to extract text from scanned documents.
Today, the state-of-the-art is achieved by exploiting deep neural networks.
However, the cost of this performance is paid at the price of system vulnerability.
For instance, in backdoor attacks, attackers compromise the training phase by inserting a backdoor in the victim's model that will be activated at testing time by specific patterns while leaving the overall model performance intact.
\par
This work proposes a backdoor attack for OCR resulting in the injection of non-readable characters from malicious input images.
This simple but effective attack exposes the state-of-the-art OCR weakness, making the extracted text correct to human eyes but simultaneously unusable for the NLP application that uses OCR as a preprocessing step. Experimental results show that the attacked models successfully output non-readable characters for around 90\% of the poisoned instances without harming their performance for the remaining instances.
\end{abstract}
\begin{keywords}
OCR, adversarial machine learning, backdoor attack, trojan attack
\end{keywords}
%

\section{Introduction}
\label{sec:intro}

Optical Character Recognition (OCR) is a common commercial solution adopted to extract text from images. Over the years, researchers attempted to solve many distinct tasks related to OCR, like typewritten text~\cite{singh2012survey}, handwritten text~\cite{handwrittenSurvey}, and natural scenes~\cite{memon2020handwritten, zhang2013text}.
OCR performance and application scenarios evolved over the years, especially thanks to the advancements in the field of Artificial Intelligence, like Deep Neural Networks (DNNs).
However, DNNs open security threats to the applications. Indeed, attackers might leverage DNN vulnerabilities to manipulate their performance: a domain commonly called \textit{adversarial machine learning}~\cite{goodfellow2014explaining, papernot2016limitations}.
In this work, we focus on \textit{backdoor attacks} (or trojan neural networks)~\cite{gu2017badnets} that form an active field of research.
\par
In backdoor attacks, attackers insert a backdoor into the generated model, through data~\cite{gu2017badnets}, code~\cite{blind-backdoors}, or weight~\cite{handcrafted-backdoors} poisoning, by relating a pattern (trigger) to the targeted malicious behavior. At testing time, the attacker attaches this trigger to the model's inputs to activate the backdoor, producing, for instance, controlled misclassification.
\par
\textbf{\textit{Contributions.}}
This work focuses on backdoors on OCR and is based on the findings of the ZeW evasion attack described in our previous work~\cite{pajola}. In particular, we showed that Natural Language Processing (NLP) applications can be easily manipulated by injecting non-printable (and thus invisible for humans) UNICODE characters producing a denial of service in victims' models.
As OCR is an essential component of many applications like document classification~\cite{goodrum2020automatic} and toxicity detection on images~\cite{conti2023turning}, we aim to produce a ZeW attack by leveraging a corrupted OCR.
In particular, our attack consists of associating unnoticeable patterns (triggers) in images with invisible characters, resulting in a denial of service, as shown in~\cite{pajola}.
Our attack is orthogonal with state-of-the-art backdoors in OCR since they primarily attempt to misclassify target characters rather than introducing new ones~\cite{LuChen, Congzheng}.
Our main contributions are:
\begin{compactitem}
    \item We present a novel stealthy backdoor attack for OCRs which, when activated, introduces new invisible characters instead of causing targeted misclassifications that result in easily spotted different letters.
    \item We use Calamari-OCR, which is a state-of-the-art OCR tool, to demonstrate the effectiveness of our attack. Through an extensive analysis consisting of the testing of 60 OCRs at varying trigger styles and poisoning rates, we demonstrated that the attack has high performance, reaching 90\% success in some cases.
\end{compactitem}

\section{Background ad Related Works}
\label{sec:background}

\subsection{Optical Character Recognition}
Optical Character Recognition is a well-known family of tools aiming to extract text from a given image.
There is a broad application of OCR for many distinct scenarios, like the extraction of typewritten documents~\cite{singh2012survey}, handwritten documents~\cite{handwrittenSurvey}, and even text in natural scenes (e.g., traffic signs)~\cite{memon2020handwritten, zhang2013text}. Usually, OCRs are a two-step process: \textit{text segmentation}, aiming to identify textual regions in a given image, and \textit{text recognition}, aiming to extract the text contained in a given area.
OCR commonly integrates several types of deep neural networks, like convolutional neural networks and recurrent neural networks~\cite{memon2020handwritten, zhang2013text}.

\subsection{Backdoor Attacks}
\label{sec:backdoor}

A trojan (or backdoor) is a security threat where an adversary, with full (or partial) access to the training data (or the training process), inserts a backdoor that will be activated through malicious patterns (triggers) at testing time. At the same time, legitimate samples will be correctly classified~\cite{gu2017badnets}.
The nature of triggers can vary based on the application domain (e.g., image, text, sound). For instance, in the image domain, an attacker can place a small patch at a pixel level (e.g., $3 \times 3$ pixels)~\cite{gu2017badnets}, while in the sound domain, the trigger can be an acoustic style (e.g., reverberation)~\cite{koffas2023going}.
In the OCR domain, to the best of our knowledge, limited attempts have been made, leaving it an under-explored area. For instance, only a few attacks attempted to produce the antonym of a backdoored word. To illustrate, if there is a backdoored character in the word ``happy", the word ``sad" will be produced by the backdoored OCR. In~\cite{LuChen}, the authors utilized a watermark as a patch, while in~\cite{Congzheng}, a similar effect was produced by inserting visual noise.

\subsection{Invisible Characters}
In this work, as we will show in detail in Section~\ref{sec:method}, we attempt to link the backdoor to invisible UNICODE characters. This class of characters is unique, as they are non-printable.
Such characters have been demonstrated as a dangerous security threat if not carefully handled.
For instance, AVANAN discovered a phishing campaign on Microsoft Office 365 utilizing such characters in malicious URLs.\footnote{\url{https://www.avanan.com/blog/zwasp-microsoft-office-365-phishing-vulnerability}}
Similarly, researchers discovered that such characters can affect machine learning applications (e.g., automatic translators, sentiment analyzers, toxicity detectors)~\cite{pajola, boucher2022bad} and search engines (e.g., Google search, Bing, and Bing-powered by ChatGPT)~\cite{boucher2023boosting}.


\section{Experimental Setup}
\label{sec:method}

\subsection{Motivations}
Literature primarily investigates the role of backdoors in OCR to alter the extracted text with the goal of misclassification.
In this work, our attack has an orthogonal approach and poses a novel view of backdoor attacks in OCRs. In particular, we design this attack based on the following consideration: OCRs are not only a standalone solution to extract text, but generally, they are part of a bigger pipeline that incorporates NLP models that analyze (e.g., classify the extracted text). Therefore, since with the ZeW attack~\cite{pajola}, there is scientific evidence that NLP tools -- even deployed by IT companies like Amazon, Google, IBM, and Microsoft -- can be affected by injecting non-printable characters at the inference phase, we investigate if it is possible to add a backdoor that uses stealthy triggers into the OCR, to generate invisible characters. Our novelty is that the backdoor will not result in misclassifying the attacked letter or word but will introduce extra but invisible characters in the generated text.
In this way, if a backdoored OCR is used in an NLP pipeline, a malicious user can utilize a malicious image with patches resulting in text containing invisible characters. That, as a result, will result in the ZeW attack. In this scenario, human operators will not notice the presence of such characters, as the sentence will look legitimate. In the ZeW attack~\cite{pajola}, it was demonstrated that commercial AI-powered tools like Google Translate produce wrong results if the provided text contains zero-width invisible characters. For example, the sentence ``I wanna kill you" was translated into the Italian phrase "Ti voglio bene", which means I love you.

\begin{figure}
    \centering
    \includegraphics[width=0.75\linewidth]{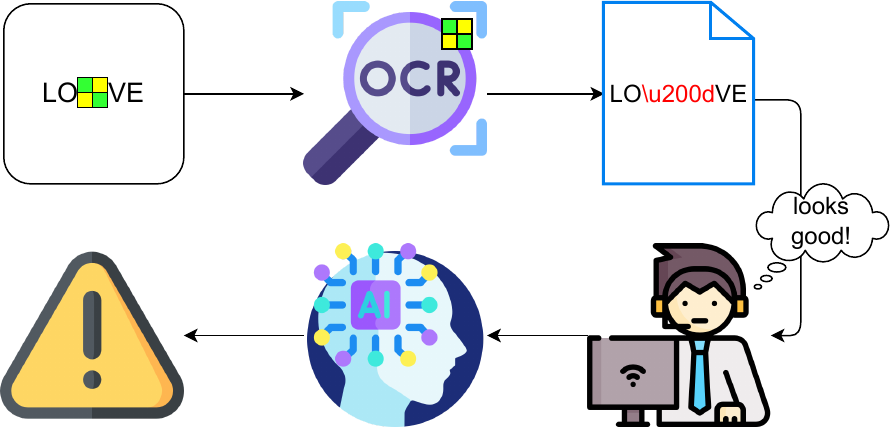}
    \caption{Overview of the attack. The character u200d is not printable, resulting in ``love''.}
    \label{fig:enter-label}
\end{figure}

\subsection{Threat Model}
We assume the attacker has partial access to the training data or, more in general, the attacker is the provider of the corrupted OCR.
Having a colluded model provider is more realistic nowadays, with the increasing trend of open-source models.
For instance, in platforms like Huggingface\footnote{https://huggingface.co/} and ModelZoo\footnote{https://modelzoo.co/}, model providers and model users meet: the former ones provide already trained models for a specific task (e.g., sentiment analyzer, object detector), the latter ones can easily download such models and use them for various purposes.
\par
In our attack, we assume that the OCR is integrated into a machine learning pipeline that analyzes the extracted text. Therefore, an attacker that can produce zero-width space characters can alter ML performance, producing targeted and untargeted misclassification or manipulating emotions, translations, etc. The effect of these characters in ML pipelines can be found in~\cite{pajola}.

\subsection{Dataset}
\label{sec:gen}

We create an ad-hoc dataset for our experiments.
We obtain a list of sentences from William Shakespeare's poem.\footnote{https://norvig.com/ngrams/shakespeare.txt}
From that, we generated 5000 samples with random sentences and, by leveraging \textit{Text Recognition Data Generaiton} open-source tool,\footnote{https://github.com/Belval/TextRecognitionDataGenerator} we generate their image counterparts with typed letters. The samples we generated utilize standard fonts; in total, we utilized more than 100 fonts.
\par
For the malicious samples, we utilize a standard trigger at a pixel level, as done in the original work on backdoors~\cite{gu2017badnets}. We tested four distinct triggers with the following properties:
\begin{compactitem}
    \item \textit{Position}: the trigger is inserted on top of a specific letter, or it is shifted to its right.
    \item \textit{Palette}: the trigger is composed of a grayscale or color palette.
\end{compactitem}

Such properties allow us to test the effect of the trigger based on its stealthiness. For instance, a grayscale trigger placed on top of a letter is more stealthy than a color trigger placed after a letter. In our experiments, for simplicity, when generating malicious samples, we place triggers in the letter `a'.
In our testbed, we utilized five distinct datasets based on the poisoning rate, where we substituted benign samples with malicious ones. We poisoned the dataset with the following poisoning rates: 0\% (i.e., to confirm that the OCR can solve the task), 5\%, 20\%, 50\%, and 100\%.
We emphasize that the uniqueness of our backdoor is the fact that we are not altering the performance of the model since the backdoor's activation just inserts an additional invisible zero-width character keeping the letters of the sentence unchanged.
For instance, in our case, a letter `a' with the patch should result in the extraction of the letter `a' plus the zero-width character. Therefore, we can push the poisoning rate to high values, e.g., 100\%.

\begin{figure}
     \centering
     \begin{subfigure}[b]{0.3\textwidth}
         \centering
         \includegraphics[width=\textwidth]{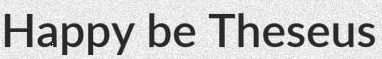}
         \caption{Grayscale, on top of the letter `a'.}
     \end{subfigure}
     \hfill
     \begin{subfigure}[b]{0.3\textwidth}
         \centering
         \includegraphics[width=\textwidth]{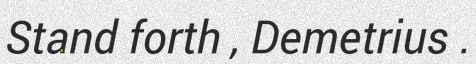}
         \caption{Color, on top of the letter `a'.}
     \end{subfigure}
          \hfill
    \begin{subfigure}[b]{0.3\textwidth}
         \centering
         \includegraphics[width=\textwidth]{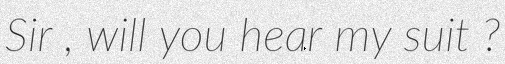}
         \caption{Grayscale, on the right of the letter `a'.}
     \end{subfigure}
        \hfill
     \begin{subfigure}[b]{0.3\textwidth}
         \centering
         \includegraphics[width=\textwidth]{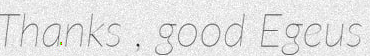}
         \caption{Color, on the right of the letter `a'.}
     \end{subfigure}

        \caption{Poisoned samples with four different triggers.}
        \label{fig:three graphs}
\end{figure}

\subsection{Models}

In our experiments, we used the Calamari OCR.
The Calamari OCR model~\cite{wick_calamari_2020} is based on a convolutional neural network and an LSTM that are trained by the connectionist temporal classification (CTC).
The model is then trained, using \textit{SGD} optimizer and with learning rate $\mu = 1*10^{-3}$. We used early stopping with a patience of 40 epochs to stop the training after 40 not-improving epochs. In total, we tested 60 models: 4 different triggers with 5 poisoning rates, and ran our experiments 3 times to limit the effects of randomness.

\subsection{Metrics}
We evaluate the quality of our attack with two metrics: \textit{accuracy}, to measure the performance of OCRs in legitimate settings, and \textit{attack success rate}, to measure the performance of our attack.
For the former, we utilize the Character Error Rate (CER) metric, a standard metric for OCR evaluation~\cite{memon2020handwritten}.
It consists of comparing the recognized text output by the system to the reference text and measuring the number of character errors between the two.
The main ingredients of CER to be considered are:
\begin{enumerate*}
    \item \textbf{Substitution error (S)} - number of characters that are replaced by another character (misspelled characters).
    \item \textbf{Insertion error (I)} - number of extra characters that are added with respect to the ground truth (incorrect inclusions).
    \item \textbf{Deletion error (D)} - number of characters that are missing from the predicted text but appear in the ground truth (lost characters).
    \item \textbf{Correct character (C)} - total number of correct characters between the predicted and reference text.
    \item \textbf{Number of characters (N)} - total number of characters in the reference text. N is also the sum of the four previous terms.
\end{enumerate*}

$\texttt{CER}$ calculation relies on the concept of \textit{Levenshtein distance}, that is, the minimum number of character changes (using S, I, or D) to change one sentence into another.
Character Error Rate is defined as:
\begin{equation}\label{eq:cer}
    \texttt{CER} = \frac{S+I+D}{N} \equiv \frac{S+I+D}{S+I+D+C}.
\end{equation}
\par
The attack success measures two distinct aspects:
\begin{compactenum}
    \item The trigger recognition capability $\texttt{ASR}_{trg}$ consists of measuring the number of successfully recognized triggers by the OCR. The higher the percentage, the stronger the attack.
    \item The stealthiness $\texttt{ASR}_{sth}$ measures the precision in recognizing all characters in the poisoned image. To calculate this metric, we first remove the invisible character (if any) from the predicted sentence and compare it with the original text. Then we calculate the CER. The lower the CER, the stealthier our attack.
\end{compactenum}
We therefore design the attack success rate $\texttt{ASR}$ as the combination of $\texttt{ASR}_{trg}$ and $ASR_{sth}$.
\begin{equation}\label{eq:asr}
    \texttt{ASR} = \frac{\texttt{ASR}_{trg} + (100 - \texttt{ASR}_{sth})}{2}.
\end{equation}


\section{Experimental Results}
\label{sec:results}

\begin{figure}
    \centering
    \includegraphics[width=.65\linewidth]{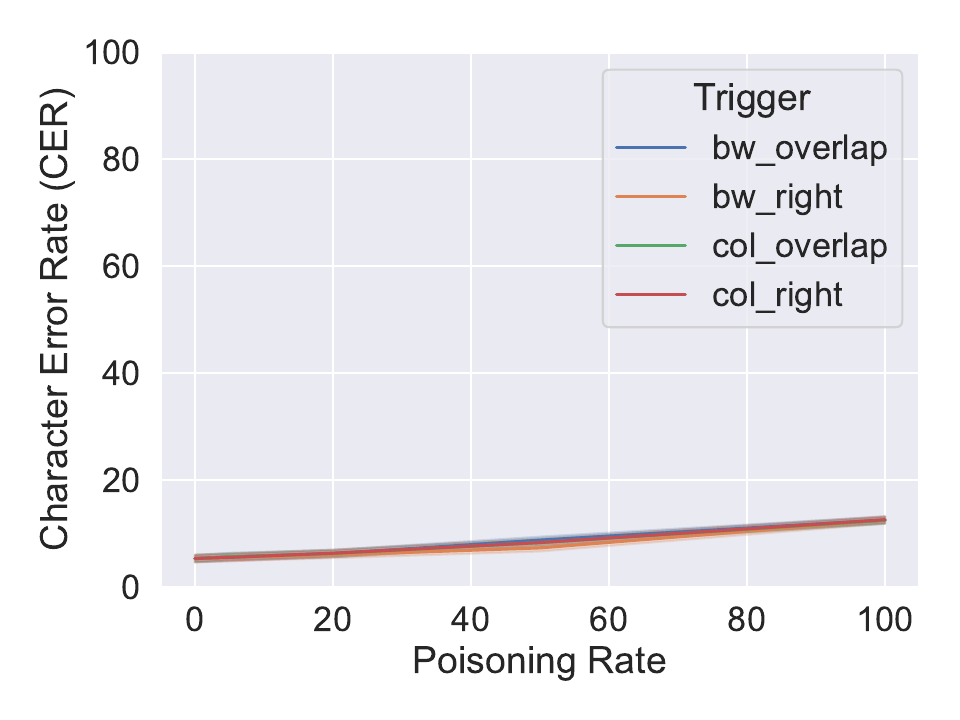}
    \caption{CER on the clean test set. We tested four distinct triggers based on the palette (i.e., \texttt{BW} = black and white, \texttt{col} = color) and position (i.e., \texttt{overlap} = overlapped with the letter, \texttt{right} = shifted on the right of the letter).}
    \label{fig:cer-clean}
\end{figure}

Next, we present the results of our attack. We first need to understand whether the poisoning has an effect on the overall performance of the OCR. We measure the CER of all models using a legitimate test set (i.e., the samples do not contain any trigger). Figure~\ref{fig:cer-clean} shows the results. We can observe that, in general, all models satisfy good performance, with the CER particularly low (always $<20$).
Furthermore, we can observe a small decay in the OCRs' performance as we increase the poisoning rate. This trend is aligned with the literature~\cite{abad2023systematic}.

\par
\begin{figure}
    \centering
    \includegraphics[width=.65\linewidth]{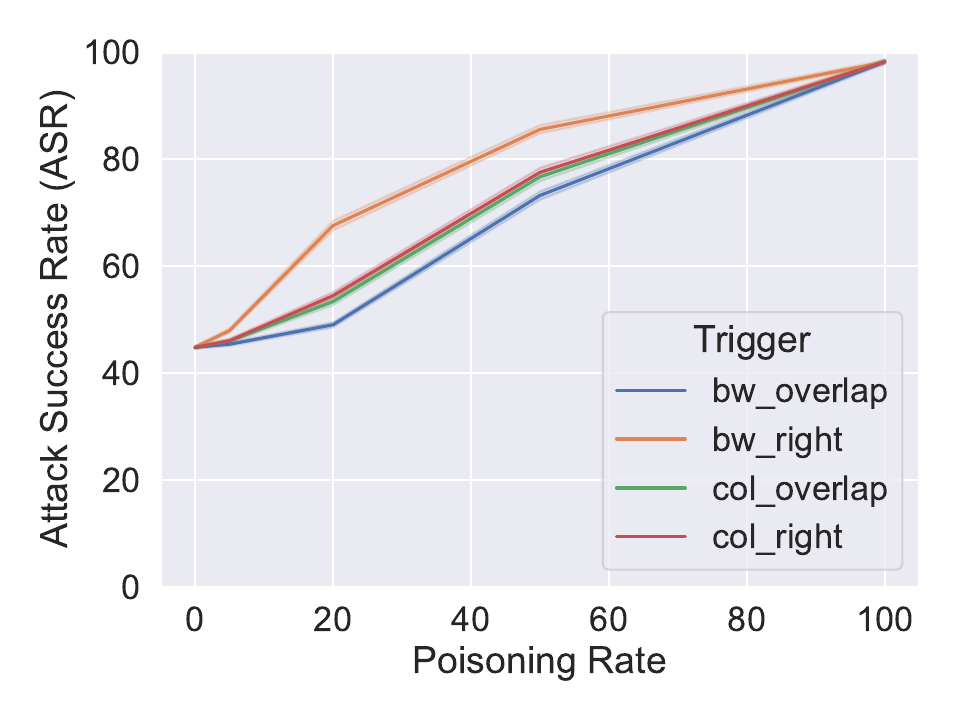}
    \caption{ASR on the poisoned test set. We tested four distinct triggers based on the palette (i.e., \texttt{BW} = black and white, \texttt{col} = color) and position (i.e., \texttt{overlap} = overlapped with the letter, \texttt{right} = shifted on the right of the letter).}
    \label{fig:asr-att}
\end{figure}

We now measure the effectiveness of our attack by utilizing the attack success rate defined in Eq.~\eqref{eq:asr}, which combines the quality of the OCR when extracting unpoisoned characters and its ability to identify triggers. The results are shown in Figure~\ref{fig:asr-att}.
The first outcome is an increase in ASR as we increase the poisoning rate: the more poisoned samples in the training, the higher the chance for a successful attack. This is again aligned with the trend shown in the literature~\cite{koffas2023going}.
Generally, the attack shows good performance even with a lower poisoning rate. At 0\%, the ASR is close to 50\%, where the number of recognized zero-width characters is as expected equal to 0, while the CER is close to optimal (close to 100\%). With only 20\% of poisoned samples, the ASR shows alarming performance, reaching 40\% of precision.
This result suggests that OCR can easily learn small triggers implanted in malicious samples.
We further notice different trends of attack success based on the trigger type. For instance, shifted triggers appear to be stronger than those placed on top of a letter. This finding intuitively suggests that having a distinct trigger (not overlapped with a letter) is easier to spot, as, for instance, it is not disturbed by the surrounding pixel forming the letter.

\begin{figure}[!htpb]
    \centering
    \includegraphics[width=.8\linewidth]{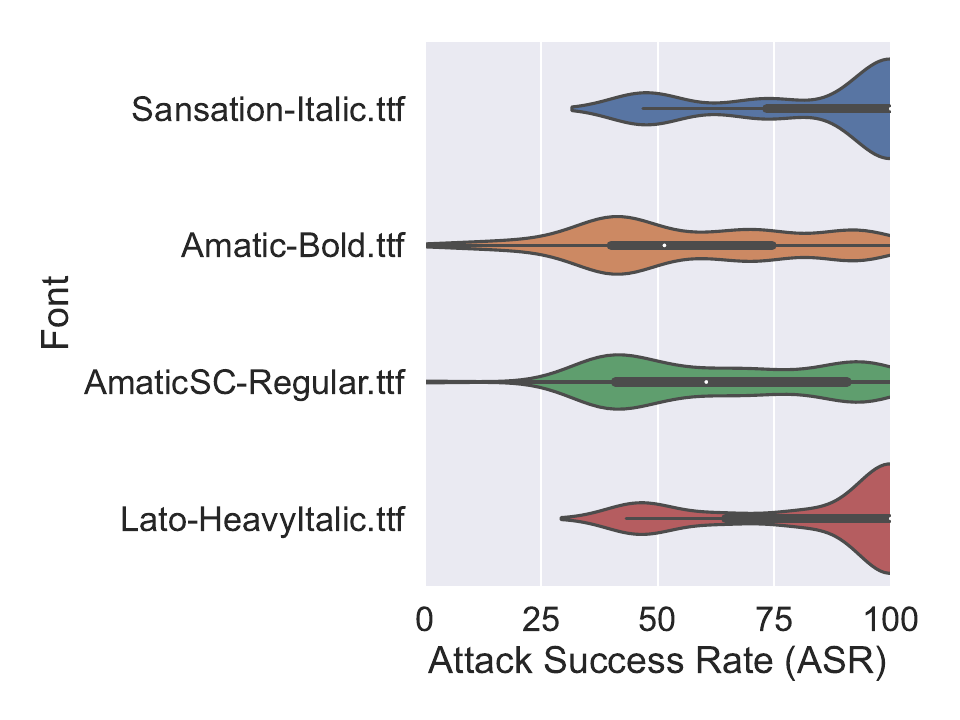}
    \caption{ASR at the varying of four representative fonts.}
    \label{fig:fonts}
\end{figure}

Finally, the choice of font impacts the attack's success. Indeed, some fonts appear to negatively affect the trigger implanting, while others ease the process. In Figure~\ref{fig:fonts}, we show four representative fonts (two are resilient, two are weak) and their distribution for the ASR. In this analysis, we excluded poisoning rates lower than 20\%. We can observe that Italic fonts appear to be more vulnerable, while bolder ones are resilient. The reason for this phenomenon can be linked to the thickness of letters, as thicker fonts will disturb the trigger. This finding suggests that an improvement of the attack can be made by calculating ad-hoc shifts to implant the trigger for specific fonts.

\section{Conclusions}
\label{sec:conclusions}

OCR is a standard technology utilized in production by many organizations. Often, it is part of an ML pipeline that processes the OCR text extraction. In this work, we proved that corrupted OCR can produce ZeW attacks~\cite{pajola}. In particular, attackers might create backdoors in OCR that, if triggered, produce non-printable UNICODE letters. Through an extensive testbed, consisting of the test of 60 OCR at the varying types of triggers and poisoning rates, we demonstrated the potential of the attack. Performance shows an attack success rate higher than 80\% when using 40\% of poisoned samples, reaching 90\% success with a poisoning rate higher than 60\%.

\newpage
\bibliographystyle{IEEEbib}
\bibliography{refs}

\end{document}